\documentclass[dvips,12pt]{article}
\input{axodraw.sty}
\usepackage{amsmath,amssymb,exscale}
\usepackage{array,multicol}
\usepackage{afterpage,float,flafter}
\usepackage{epsfig,rotating,pifont,fancybox}
\makeatletter
\@addtoreset{equation}{section}
\makeatother

\newcommand{\bea}{\begin{eqnarray}}
\newcommand{\eea}{\end{eqnarray}}

\newcommand{\mbar}{\bar{M}}
\def\simlt{\stackrel{<}{{}_\sim}}
\def\simgt{\stackrel{>}{{}_\sim}}
\newcommand{\NPB}[3]{\emph{ Nucl.~Phys.} \textbf{B#1} (#2) #3}
\newcommand{\PLB}[3]{\emph{ Phys.~Lett.} \textbf{B#1} (#2) #3}
\newcommand{\PRD}[3]{\emph{ Phys.~Rev.} \textbf{D#1} (#2) #3}
\newcommand{\PRL}[3]{\emph{ Phys.~Rev.~Lett.} \textbf{#1} (#2) #3}

\title{
\vspace*{-1.3cm}
\begin{flushright}
\normalsize{
ANL-HEP-PR-01-007 \\
EFI-2001-004  \\
FERMILAB-PUB-01/014-T}
\end{flushright}
\vspace{-2.2cm}\begin{flushleft}
\normalsize{
IFT-00/31 \\
IEM-FT-207/00\\
IFT-UAM/CSIC-00-41}
\end{flushleft}
\Large
\textbf{
Brane Effects on Extra Dimensional
Scenarios: \\
A Tale of Two Gravitons} 
\vspace*{.5cm}
\author{\large\textbf
{M.~Carena$^a$, A.~Delgado$^{b,c}$, J.~Lykken$^{a,e}$ }\\ 
\textbf{S.~Pokorski$^{a,d}$}, \textbf{M.~Quir\'os$^{a,c}$} and
\textbf{C.E.M.~Wagner$^{b,e}$}\\ \\
$^a$\normalsize\emph{Fermi National Accelerator Laboratory,
P.O. Box 500, Batavia, IL 60510, USA }\\
$^b$\normalsize\emph{HEP Division, Argonne National Laboratory,
9700 Cass Ave.,
Argonne, IL 60439, USA} \\
$^c$\normalsize\emph{Instituto de Estructura de la Materia (CSIC),
Serrano 123,
E-28006 Madrid, Spain}\\
$^d$\normalsize\emph{Institute for Theoretical
Physics, University of Warsaw, Hoza 69, 00-681 Warsaw, Poland}\\
$^e$\normalsize\emph{Enrico Fermi Institute, Univ. of Chicago, 5640
Ellis Ave., Chicago, IL 60637, USA}}}

\begin{document}
\maketitle

\begin{abstract}
We analyze the propagation of a scalar field in multidimensional theories 
which include kinetic corrections in the brane, as a prototype for 
gravitational interactions in a four dimensional brane located in a 
(nearly) flat extra dimensional bulk. We regularize the theory by 
introducing an infrared cutoff given by the size of the extra dimensions
and a physical ultraviolet cutoff of the order of the fundamental
Planck scale in the higher dimensional theory. We show that, contrary to 
recent suggestions, the radius of the extra dimensions cannot be arbitrarily 
large. Moreover, for finite radii, the gravitational effects localized on 
the brane can substantially alter the phenomenology of collider and/or
table-top gravitational experiments. This phenomenology is dictated by
the presence of a massless graviton, with standard couplings to the matter 
fields, and a massive graviton which couples to matter in a much stronger
way. While graviton KK modes lighter than the massive graviton
couple to matter in a standard way, the couplings to matter of the heavier KK
modes are strongly suppressed.
\end{abstract}

\thispagestyle{empty}
\newpage

\section{Introduction}

The existence of large extra dimensions, of size much larger than
the Planck length, has been recently proposed by Arkani-Hamed,
Dimopoulos and Dvali (ADD)~\cite{ADD} as an alternative
means of solving the hierarchy problem of the Standard Model (SM).
It has been argued that, if the SM fields were localized in a
four dimensional brane located in a compact flat spatial bulk
of radius $R$, which allows the free propagation of gravitons,
the physical Planck scale associated with the gravity properties at
distances much larger than $R$ will be given by $M_{Pl}^2 = M^{2+d} R^d$,
where $M$ is the fundamental Planck scale and $d$ is the number of
extra dimensions. For this mechanism to provide a solution of
the hierarchy problem, $M$ should be on the order of a TeV. Since
gravity will look ($4+d$) dimensional at distances much shorter than
$R$, the compactification radius cannot be larger than a millimeter.
While such a bound is inconsistent with a solution to the hierarchy
problem in the case $d = 1$, it is naturally fulfilled for
$d \geq 2$. The smaller $d$, the larger becomes $R$ and, for $d = 2$,
the predicted deviation of the gravity behavior is being
tested in table-top experiments~\cite{mm,mm2}.

Since the fundamental scale $M$ is of order a TeV, non-trivial
modifications will also appear in the ultraviolet
regime of energies close to $M$. From a four dimensional point of
view, the ultraviolet effects will be associated with the emission
of Kaluza-Klein (KK) states, representing the propagation of the graviton
in the extra dimensional bulk. Although individual KK gravitons will
couple very weakly to SM particles, the cumulative effect of KK
gravitons will lead to interactions which become strong at energy scales
of order $M$. 
For collider center-of-mass energies approaching $M$, graviton emission
can be observed as missing energy signatures, or as virtual effects that
interfere with SM processes~\cite{colliders}. Nonobservation of these
collider effects places a lower bound on $M$ of about a TeV.

From the above considerations, it seems clear that, for a compact {\it flat}
bulk space, $R$ cannot be larger than a millimeter without producing gross
deviations from four dimensional gravity at macroscopic scales.
This point of view has been recently challenged by the authors
of Ref.~\cite{porrati}.  They show that, if $d=1$ and $R$ is taken
to be infinitely large, the possible inclusion of
a local four dimensional Einstein term in the 
brane (see also~\cite{brane5}) may lead
to short--distance behavior that resembles the usual four
dimensional propagation of gravitons. 
Moreover, in Ref.~\cite{dvali} it was argued that
if the number of dimensions is larger than five, $d \geq 2$,
one obtains four dimensional behavior for the graviton at all scales.
However, for $d \geq 2$ this result is derived from expressions involving 
an ultraviolet divergent
contribution coming from the propagation of gravitons in the bulk,
and therefore the case $d \geq 2$
deserves a more detailed analysis. This is particularly so since
we are considering a non-renormalizable theory with a physical
ultraviolet cutoff which cannot be larger than the ($4+d$) dimensional
Planck scale $M$, at which gravity interactions in the extra dimensional
bulk become strong.

In this paper we will introduce both an infrared (IR) and an ultraviolet (UV)
regularization.  The IR regularization is implemented
by means of a $d$ dimensional torus with common radius $R$.
We introduce 
a physical UV cutoff $\Lambda$  by truncating the
number of KK-modes at $|\vec{n}| = \Lambda R/\pi$, where
$\vec{n}^2=n_1^2+\cdots n_d^2$.
The case of $(4+d)$ dimensional Minkowski space
considered in Refs.~\cite{porrati,dvali} will appear as the $R\to\infty$
limit. In this limit we will reproduce the results of
Ref.~\cite{porrati} for the case of one extra dimension; moreover for
$d \geq 2$ we 
reproduce the
results of Ref.~\cite{dvali} in the limit $R\to\infty$ and
$\Lambda\to\infty$.
However, when we fix the UV cutoff to a natural physical 
value $\mathcal{O}(M)$, the results we obtain are different from those 
obtained in~\cite{dvali}.

For the case of finite radius (toroidal) extra dimensions
the presence of brane gravitational
corrections could substantially modify the usual ADD scenario. In fact
we shall show that, depending on the strength of those corrections,
either the collider or table-top gravitational signatures
can be substantially altered.

The outline of this paper is as follows.
In section 2 we study the structure
of the bulk graviton propagator including the brane corrections for an
arbitrary number of flat extra dimensions. In section 3 we consider the
$R\to\infty$ limit to compare with the results of Refs.~\cite{porrati,dvali}.
In section 4 we consider the case of finite radius and analyze
the physical features of the relevant
momentum (or distance) regimes.
Section 5 is devoted to the phenomenology, including collider experiments and
gravitational table-top experiments.
Finally our conclusions and
outlook are presented in section 6 and some lengthy formulae are
exhibited in appendix A.

\section{General propagator including the brane correction}

We start with a $(4+d)$ dimensional theory with coordinates
$x^I\equiv(x^\mu,y^i)$,
where $x^\mu$ are the four dimensional space-time coordinates and
$0\leq\, y^i\, \leq R$, ($i=1,\dots, d$) 
those of the compact space and assume a
3-brane embedded in the $(4+d)$ dimensional space at $y=0$.
We shall assume that the brane thickness is small compared to the
$(4+d)$ dimensional Planck length, and model it by a delta
function in the extra dimensions.
Following Refs.~\cite{porrati,dvali}, we shall
consider, for simplicity, a single bulk scalar field $\phi (x^I)$,
instead of the
more complicated tensor structure of a graviton,
propagating in the $(4+d)$ dimensional
space, with the action:
\begin{equation}
\label{action}
S_{4+d}=\int d^4x d^dy\left\{M^{2+d} \partial_I\phi(x,y)\partial^I\phi(x,y)
+\mbar^2\delta(y)
\partial_\mu\phi(x,0)\partial^\mu\phi(x,0)\right\}
\end{equation}
where $M$ is the higher dimensional Planck scale,
$\mbar$ is the coefficient of the brane-generated correction,
and we have scaled the $\phi$ field to be dimensionless.
Eq.~(\ref{action}) should be regarded as the leading terms in a
bulk+brane effective action for dynamics in a $(4+d)$ dimensional
flat space background (we are ignoring bulk+brane cosmological
constant terms).
The effective action will be valid up to
some physical ultraviolet cutoff $\Lambda$, where it matches onto
a more fundamental UV description.
Since bulk gravity becomes strongly interacting at the energy scale
$M$, we expect that the case of interest is
$\Lambda$ not much larger than $M$.

The brane-generated kinetic term with coefficient $\mbar$ could be 
 induced from
the coupling between bulk gravity
(the bulk scalar, in our simplified treatment)
and the brane matter fields. As noted in \cite{porrati,dvali},
graviton vacuum polarization diagrams with brane matter loops
will generically produce such a term. For example, if there are
$N$ heavy brane particles of mass $m$, they will generate a contribution
to $\mbar$ on the order of $Nm^2$ times a loop
factor~\cite{brane1}-\cite{brane4}. Even in an effective theory with
a cutoff $\Lambda \simeq M$, it is consistent to imagine that $\mbar$
may be much larger than $M$. This occurs in the example above if $N$
is large; it could also arise from dynamical sources such as coupling
to brane or bulk fields which have large vacuum expectation values.
Other possibilities include a large $\mbar$ related to physics of the
fundamental UV theory.
On the other hand $\mbar\gg M$ introduces a hierarchy problem,
which would have to be addressed in a complete model. 

We are interested in obtaining
the Green function for the propagation of the scalar field
from a point in the brane to any point in the $(4+d)$ dimensional
space. The equation for the corresponding Green
function, after making a Fourier transform over the four dimensional
space-time coordinates,
is
\begin{equation}
\label{generaleq}
[M^{2+d}(p^2-\Delta_d)+\mbar^2p^2\delta^{(d)}(y_i)]G_d(p,y_i)=
\delta^{(d)}(y_i)
\end{equation}
where $p^2=p_1^2+p_2^2+p_3^2+p_4^2$ is the four dimensional euclidean momentum
and $\Delta_d$ is the laplacian operator in $d$ dimensions.

The solution to Eq.~(\ref{generaleq}) can be written as follows, 
\cite{porrati,dvali}:

\begin{equation}
G_d(p,y_i)=\frac{D_d(p,y_i)}{M^{2+d}+\mbar^2 p^2 D_d(p,0)}
\label{solution1}
\end{equation}
\noindent
where $D_d(p,y_i)$ is the solution to the equation
\begin{equation}
(p^2-\Delta_d)D_d(p,y_i)=\delta^{(d)}(y_i)\ .
\label{laplace}
\end{equation}

A way to solve Eq.~(\ref{laplace}) is by Fourier series,
since the extra dimensions
are finite. We
will only be interested in the behavior of the
higher dimensional propagator on the 3-brane,
that is for $y_i=0$. In this case the form for $D_d(p,0)$
is just the sum
\begin{equation}
D_d(p,0)=\frac{1}{R^d}\sum_{\vec{n}}
\frac{1}{p^2+\frac{\vec{n}^2 \pi^2}{R^2}}
\label{KK}
\end{equation}
where the vector $\vec{n}$ is defined as, $\vec{n}=(n_1,\dots,n_d)$.

The Green function on the 3-brane $G_d(p,0)$ could also be obtained after
integration of the action (\ref{action}) over the extra dimensional
coordinates $y^i$, which yields the four dimensional Lagrangian for the
KK-states $\phi^{(\vec{n})}$,
\begin{equation}
\label{lag4}
\mathcal{L}_4=M^{2+d}\,R^d \sum_{\vec{n}}\left( \partial_\mu \phi^{(\vec{n})}
\partial^\mu \phi^{(\vec{n})}-
\frac{\vec{n}^2}{R^2}\,\phi^{(\vec{n})}\phi^{(\vec{n})}\right)
+\mbar^2\sum_{\vec{n},\,\vec{m}}\partial_\mu \phi^{(\vec{n})}
\partial^\mu \phi^{(\vec{m})}\ .
\end{equation}

The Green function on the 3-brane
is the sum over four dimensional propagators. To compute it
we consider the first two
terms in (\ref{lag4}) as the unperturbed Lagrangian, giving rise to the
sum
\begin{center}%
${\displaystyle D_d(p,0)/M^{2+d}=\sum_{\vec{n}}\quad (\vec{n})}$
\SetScale{0.6}
\begin{picture}(70,30)(0,28)\hspace{-.5cm}
\Photon(20,50)(90,50){4}{10}
\end{picture}\hspace{-.8cm}$(\vec{n})\equiv$
\begin{picture}(70,30)(0,28)\hspace{-.5cm}
\Photon(20,50)(90,50){4}{10}
\end{picture}
\end{center}
and the last term as the perturbed Lagrangian,
giving rise to the mixing
\begin{center}%
$-\, p^2\, \mbar^2\equiv\quad (\vec{n})$
\SetScale{0.6}
\begin{picture}(70,30)(0,28)\hspace{-.5cm}
\Photon(20,50)(90,50){4}{10}
\Vertex(90,50){5}
\Photon(95,50)(165,50){4}{10}
\end{picture}\hspace{.7cm}$ (\vec{m})$
\end{center}
In this way the complete Green function $G_d(p,0)$ can be computed as
\begin{center}
$G_d(p,0)\equiv$
\SetScale{0.6}
\begin{picture}(50,30)(0,28)\hspace{-.5cm}
\Photon(20,50)(80,50){4}{8}
\end{picture}
\hspace{-.65cm}
+
\SetScale{0.6}
\begin{picture}(50,30)(0,28)\hspace{-.5cm}
\Photon(20,50)(80,50){4}{8}
\Vertex(80,50){5}
\Photon(85,50)(145,50){4}{8}
\end{picture}\hspace{1.5cm}
\hspace{-.65cm}
+
\SetScale{0.6}
\begin{picture}(50,30)(0,28)\hspace{-.5cm}
\Photon(20,50)(80,50){4}{8}
\Vertex(80,50){5}
\Photon(85,50)(145,50){4}{8}
\Vertex(145,50){5}
\Photon(150,50)(210,50){4}{8}
\end{picture}$\hspace{2.5cm}+\quad\cdots$
\end{center}

\noindent and coincides with the expression obtained from (\ref{solution1})
and (\ref{KK}).

We have computed the sum over the four dimensional propagators in the
original field basis, without explicitly diagonalizing the lagrangian
(\ref{lag4}). In the basis of canonically normalized mass eigenstates
for the lagrangian (\ref{lag4}), $G_d(p,0)$ is just the sum over
the exact KK propagators, with mode dependent residues proportional
to the square of the coupling of the corresponding eigenstates  to matter,
whereas $D_d(p,0)$ is the sum over the unperturbed KK propagators.

In principle the summation is over all KK modes, but since the action
(\ref{action}) is that of an effective
theory we should cutoff the sum when the mass of the modes are of order
of the scale $\Lambda$, where $\Lambda$ is the cutoff of the theory.
That is we will sum up to a maximum $\vec{n}^2_{\rm max}=(\Lambda R/\pi)^2$.
We expect the cutoff $\Lambda$ to be of the order of the characteristic 
scale of the
underlying higher dimensional theory.

For large values of $R$, as those that we will be interested in this paper,
we can approximate the summation in (\ref{KK})
by an integral over the $n_i$. After subtracting the zero-mode we can write
(\ref{KK}) as,
\begin{equation}
D_d(p,0)=\frac{1}{p ^2R^d}\left(1+\Omega_d \frac{R^d p^d}{\pi^d}
\int_{\frac{\pi}{Rp}}^{\frac{\Lambda}{p}}\frac{x^{d-1}dx}{1+x^2}\right)
\label{integral1}
\end{equation}
where $\Omega_d=\frac{2\pi^{d/2}}{2^d \Gamma[\frac{d}{2}]}$ is the
solid angle sustained in $d$ dimensions for $y_i > 0$.
This expression can be considered as the $y_i\to 0$ limit of
the bulk propagator

\begin{equation}
D_d(p,y)=\frac{1}{p ^2R^d}\left(1+\left(\frac{\pi}{2}\right)^{\frac{d}{2}}
\frac{R^d p^d}{\pi^d}
\int_{\frac{\pi}{Rp}}^{\frac{\Lambda}{p}}\frac{x^{d-1}
\; J_{d/2-1}(x p y)
\; dx}{(xpy)^{d/2-1}(1+x^2)}\right)
\label{integral1y}
\end{equation}
where the $J_{d/2-1}$ are Bessel functions and $y\equiv|\vec{y}|$.

The explicit expressions for the Green functions $D_d(p,0)$ are given in
Appendix A, Eqs.~(\ref{odd}) and (\ref{even}), as well as their analytic
continuation into the Minkowski space-time, $D_d(s)$,
Eqs.~(\ref{oddM}) and (\ref{evenM}). For the purpose of this paper we will
find it more convenient to work in Minkowski space-time so we will use, from
here on,` the Green functions $D_d(s)$, as given by
(\ref{oddM}) and (\ref{evenM}) as well as the corresponding $G_d(s)$ functions
as defined by
\begin{equation}
G_d(s)=\frac{D_d(s)}{M^{2+d}-\mbar^2 s D_d(s)}\ .
\label{solutionM}
\end{equation}

The cases where $d=1$, $2$ are particularly
interesting, and they will be analyzed separately in the following sections.
The corresponding Green functions, within the approximation given
in Eq.~(\ref{integral1}), can be written as:
\begin{equation}
G_1(s)={\displaystyle
\frac{-1+\frac{R\sqrt{s}}{\pi}
\left[i\, \frac{\pi}{2}-\tanh^{-1}\left(\frac{\sqrt{s}}{\Lambda}\right)
+\tanh^{-1}\left(\frac{\pi}{R\sqrt{s}}\right)\right]}
{\left(M^3R+\mbar^2\right)s-
\frac{\mbar^2 R\, s\sqrt{s}}{\pi}
\left[i\, \frac{\pi}{2}-\tanh^{-1}\left(\frac{\sqrt{s}}{\Lambda}\right)
+\tanh^{-1}\left(\frac{\pi}{R\sqrt{s}}\right)\right]} }
\label{5d}
\end{equation}
and
\begin{equation}
G_2(s)={\displaystyle
\frac{-1+\frac{R^2\, s}{2\pi}\,\left[i\, \frac{\pi}{2}+\frac{1}{2}
\log\left(\frac{\Lambda^2}{s}-1\right)-\frac{1}{2}
\log\left(1-\frac{\pi^2}{R^2s}\right)\right] }
{\left(M^4R^2+\mbar^2\right)s-
\frac{\mbar^2R^2 s^2}{2\,\pi}\, \left[i\, \frac{\pi}{2}+\frac{1}{2}
\log\left(\frac{\Lambda^2}{s}-1\right)-\frac{1}{2}
\log\left(1-\frac{\pi^2}{R^2s}\right)\right] } }\ .
\label{6d}
\end{equation}

We are now ready to study the behavior of the Green functions in the different
regimes of values for $R$ and, in particular, the impact on them of the
presence of the brane corrections proportional to $\mbar$.

\section{The limit of infinite size extra dimensions}

In this section we will analyze the behavior of the theory for extremely
large values of the radius $R$ and, in particular, its behavior in the
$R\to\infty$ limit, that has been recently proposed as an alternative way of
localizing gravity in the 3-brane. We will make a separate analysis of the
five and six dimensional cases and then we will analyze the general case for
$d>2$.

\subsection{One extra dimension}

The Green function (\ref{5d}) in the $R\to\infty$ limit is given by,
\begin{equation}
\label{1large}
G_1(s)\simeq \frac{i\, \frac{\pi}{2}- \tanh^{-1}
\left(\frac{\sqrt{s}}{\Lambda}\right)}
{\pi\, M^3\sqrt{s}-\mbar^2\, s \left[\frac{i\, \pi}{2}- \tanh^{-1}
\left(\frac{\sqrt{s}}{\Lambda}\right)\right] }\ .
\end{equation}
In the region ($\sqrt{s}\ll \Lambda$), $\tanh^{-1}(\sqrt{s}/\Lambda)
\simeq \sqrt{s}/\Lambda$
and the Green function (\ref{1large}) behaves approximately as,
\begin{equation}
\label{1IR}
G_1(s)\simeq -\frac{1}{i\,2\, M^3\, \sqrt{s}+\mbar^2\, s}\ .
\end{equation}

We reproduce in this way the propagator found in Ref.~\cite{porrati} for the
case of one infinite flat extra dimension. The physics described by (\ref{1IR})
was already analyzed in Ref.~\cite{porrati}. For distances $r\gg r_c$, where
the critical distance is given by $r_c\simeq \mbar^2/2\, M^3$ the linear term
($\sqrt{s}$-term)
in (\ref{1IR}) dominates and the propagator behaves as a five dimensional one. 
However, as
already observed in Ref.~\cite{porrati}, even for the most favorable
case of $\mbar\simeq M_{Pl}$ and $M\simeq 1$ TeV, the critical
distance  is not large enough, $r_c \simeq 10^{15}$ cm, and enters in
conflict with well tested Newtonian predictions.
Moreover, for $r\ll r_c$ the quadratic term in (\ref{1IR}) (corresponding to a
four dimensional theory with a propagator $\sim -1/\mbar^2 s$),
dominates but the theory is described by a scalar-tensor
theory of gravity, with an additional scalar attractive force corresponding to
the five degrees of freedom of a 4D massive or 5D massless graviton.

\subsection{Two extra dimensions}

A similar analysis can be done for the $d=2$ case using the Green function
(\ref{6d}). The $R\to\infty$ limit of (\ref{6d}) is,
\begin{equation}
\label{2large}
G_2(s)\simeq \frac{i\,\pi+\log\left(\Lambda^2/s\right)}
{4\pi\, M^4-\mbar^2\, s\left[i\,\pi+\log\left(\Lambda^2/s\right)\right]}\ .
\end{equation}

Should we take the limit of (\ref{2large}) when $\Lambda\to\infty$ we would
obtain for the Green function the behavior $G_2(s)\sim- 1/\mbar^2 s$,
which, in the full gravity theory, would correspond to the graviton propagator
in a  four dimensional tensor theory of gravity,
reproducing the result in Ref.~\cite{dvali}. However since,
as we mentioned in the previous section,
$\Lambda=\mathcal{O}(M)$, in the limit $s\ll M^2$ we cannot neglect the
constant term in the denominator of (\ref{2large}) for all momenta.
As we did in the five dimensional case, we
can compute the critical distance
on the brane $r_c$ such that for $r\gg r_c$ the constant term dominates in the
denominator of the Green function (\ref{2large}). In this region the Green
function is logarithmic and as such it corresponds to the propagator in 
a six dimensional
theory. Only for distances $r\ll r_c$ the quadratic term dominates,
leading to the typical behaviour of the propagator in a
four dimensional theory.

The value of $r_c$ cannot be given analytically, as in the previous case.
However a good approximation is given by,
\begin{equation}
\label{rc6}
r_c\simeq \frac{\mbar}{M^2}\ \sqrt{\frac{1}{2\,\pi}\,\log\frac{\mbar}{M}}
\equiv \frac{1}{\mathcal{M}_2}\ ,
\end{equation}
where we have use that $\Lambda \simeq M$.
On the other hand, for values of $s$ close to $\mathcal{M}_2$ 
the Green function (\ref{2large}) can be approximated by,
\begin{equation}
\label{2large2}
G_2(s)\simeq -\frac{1}{\mbar^2}\
\frac{1+i\gamma_2}
{s-\mathcal{M}_2^2+i s\;\gamma_2}
\end{equation}
which describes the propagation of a resonance of mass $\mathcal{M}_2$ given by
(\ref{rc6}) and width governed by
\begin{equation}
\label{gamma2}
\gamma_2=\frac{\pi}{2}\;\frac{1}{\log\,\mbar/M}\ .
\end{equation}

Using the previous values, $\mbar \simeq M_{Pl} $ and $M \simeq 1$
TeV, we obtain from Eq.~(\ref{rc6})
$r_c\simeq 5 $ mm, which corresponds to
$\mathcal{M}_2\simeq 5\times 10^{-5}$ eV.
This low value of $r_c$ is ruled out since it would imply
that the gravity propagator behaves as six dimensional 
for $r > r_c$ and we know that for distances larger 
than sub-millimeter~\cite{mm}
gravity interactions are well described by
four dimensional Einstein gravity. 

\subsection{More than two extra dimensions}

For more than two extra dimensions the analysis can be done in full generality
using Eqs.~(\ref{oddM}) and (\ref{evenM}).
The general propagator for $R\to\infty$ is, for $d$ odd,
\begin{equation}
D_d(s)=\frac{\Omega_d}{\pi^d}\,s^{\frac{d}{2}-1}
\left[i\,\frac{\pi}{2}-\tanh^{-1}\left(\frac{\sqrt{s}}{\Lambda}\right)
+
\sum_{n=1}^{(d-1)/2}\frac{1}{d-2n}\left(
\frac{\Lambda^2}{s}\right)^{\frac{d}{2}-n}\right]
\label{oddinfty}
\end{equation}
and for $d$ even:
\begin{equation}
D_d(s)=\frac{\Omega_d}{\pi^d}\,s^{\frac{d}{2}-1}
\left[i\,\frac{\pi}{2}+\frac{1}{2}\log\left(\frac{\Lambda^2}{s}-1\right)
+
\sum_{n=1}^{\frac{d}{2}-1}\frac{1}{d-2n}\left(
\frac{\Lambda^2}{s}\right)^{\frac{d}{2}-n}\right]\ .
\label{eveninfty}
\end{equation}

In the regime $s\ll \Lambda^2$ both formulae behave in the same way:
\begin{equation}
D_d(s)\simeq \omega_d\,\left[i\, \frac{\pi}{2}(d-2)\, s^{\frac{d}{2}-1}
+\Lambda^{d-2}\right]
\label{aproxd}
\end{equation}
where $\omega_d$ for $ d > 2$ is given by
\begin{equation}
\label{omegad}
\omega_d=\frac{\Omega_d}{(d-2)\, \pi^d}\ . 
\end{equation}
Inserting the above equation into the full propagator we find:
\begin{equation}
G_d(s)\simeq -\frac{1}{\mbar^2}\;\frac{1+i\, \frac{\pi}{2}(d-2)
\left(\sqrt{s}/\Lambda\right)^{d-2} }
{s-\mathcal{M}_d^2+i\, \frac{\pi}{2}(d-2)s\,
\left(\sqrt{s}/\Lambda\right)^{d-2} }
\label{fullaprox}
\end{equation}
where $\mathcal{M}_d$ is defined by
\begin{equation}
\label{masad}
\mathcal{M}_d =
\frac{1}{\sqrt{\omega_d}}
\ \frac{M^2}{\mbar} 
\left(M/\Lambda\right)^{d-2} \simeq
\frac{1}{\sqrt{\omega_d}}
\ \frac{M^2}{\mbar}\  .
\end{equation}
where the last expression is valid for $\Lambda \simeq M$.

Again, as in the previous subsection, in the limit $\Lambda\to\infty$ we would
obtain for the Green function the behavior $G_2(s)\sim- 1/\mbar^2 s$,
corresponding to the propagator in a 
four dimensional tensor theory of gravitation,
reproducing the results of Ref.~\cite{dvali}. However, since
$\Lambda=\mathcal{O}(M)$, what we obtain from (\ref{fullaprox})
corresponds to a massive resonance with a mass $\mathcal{M}_d$, width
controlled by the function
\begin{equation}
\label{width}
\gamma_d(s)=\frac{\pi}{2}(d-2)\left(\frac{\sqrt{s}}{M}\right)^{d-2}
\end{equation}
and propagator given by
\begin{equation}
\label{full}
G_d(s)\simeq -\frac{1}{\mbar^2}
\frac{s-\mathcal{M}_d^2-s\gamma_d^2-i\,\gamma_d \mathcal{M}_d^2}
{\left(s-\mathcal{M}_d^2\right)^2+s^2\gamma_d^2}\ .
\end{equation}
~\\
Observe that, since we are assuming that $s \ll M^2$, $\gamma_d(s)$
may be approximately written as
\begin{equation}
\gamma_d(s) \simeq \frac{\pi}{2} \frac{(d-2)}
{\left[\left(\frac{M}{\sqrt{s}}\right)^{d-2} - 1\right]}
\label{generalw}
\end{equation}
what allows to make connection with the two dimensional case.

Using again  $\mbar \simeq M_{Pl} $ and $M \simeq 1$ TeV,
we get from (\ref{masad}) $\mathcal{M}_d\simgt 10^{-4}$ eV.
A massive graviton as in (\ref{full}) is excluded,
which leads to ruling out the scenarios~\cite{dvali} with infinite size
extra dimensions in flat space.

Thus, from the four dimensional point view, for $d>1$ the theory has a
mode, whose mass has a non-vanishing value in the $R\to\infty$
limit. As follows from Eq.~(\ref{1IR}),
such a mode is absent for $d=1$, what is also consistent with the
results in the case of finite size extra dimensions discussed in
section 4.

\section{The case of finite size extra dimensions}

In this section we will consider the modification of the ADD scenario by
the effect of brane gravitational corrections.
We shall describe the Green function in the region
where $\sqrt{s}\ll \Lambda$, and we shall put emphasis on the case in
which the cutoff of the effective theory
$\Lambda$ is identified with 
the higher dimensional Planck
scale $M$. The physical (four dimensional) Planck mass is
defined as
\begin{equation}
\label{mplanck}
M_{Pl}^2=M^{d+2} R^d+\mbar^2\ ,
\end{equation}
and it is associated with the interaction strength of the massless 
graviton appearing in the spectrum of the theory for any value of $d$.
We shall always assume that $\mbar^2$
is small compared to the first
term in Eq. (\ref{mplanck}), which proceeds
from the dimensional reduction of the higher dimensional theory\footnote{
In the case  $\mbar^2 \gg M^{d+2} R^d$  
the theory behaves like an ordinary theory of
gravity in four dimensions, apart from very weak corrections, which
become weaker the closer $\mbar$ is to the physical Planck scale $M_{Pl}$,
Eq. (\ref{mplanck}).}.
We will be subsequently interested in the 
IR ($R\sqrt{s}\ll 1$) and UV ($R \sqrt{s}\gg 1$) regions.

\subsection{The IR region}

In the IR region, $R\sqrt{s}\ll 1$, the Green function $D_d(s)$ can be written as,
\begin{equation}
\label{DIR}
D_d(s)\simeq -\frac{1}{R^d\, s}+\frac{\Omega_d}{(d-2)\pi^2 R^{d-2}}
\left[\left(\frac{\Lambda \; R}{\pi}\right)^{d-2}-1\right]
\end{equation}
which is valid for any dimension $d$~\footnote{The case $d=2$ should be taken
from (\ref{DIR}) as a limit.}.

The Green function $G_d(s)$ behaves then as a massless
particle with gravitational coupling given by $1/M_{Pl}^2$, which corresponds
to Einstein gravity, plus a massive particle with coupling $\sim 1/\mbar^2$.
In fact, we can write (\ref{solutionM}) as:
\begin{equation}
\label{GIRpoles}
G_d(s)\simeq -\frac{1}{M_{Pl}^2\, s}-
\frac{1}{\mbar^2}\left(1-\frac{\mbar^2}{M_{Pl}^2}\right)\,
\frac{1}{s-\mathcal{M}_d^2}
\end{equation}
with the mass $\mathcal{M}_d$, for $d\geq 2$,  given by,
\begin{equation}
\label{mIR}
\mathcal{M}_d^2= 
\left(\frac{M}{\Lambda}\right)^{d-2} \;
\frac{M^2}{\omega_d}\left(
\frac{M^2}{\mbar^2}+\frac{1}{(RM)^d}\right)=
\left(\frac{M}{\Lambda}\right)^{d-2} \;
\frac{1}{\omega_d}\ \left(1-\frac{\mbar^2}{M_{Pl}^2}\right)^{-1}
\,\frac{M^2}{\mbar^2}\, M^2\ .
\end{equation}
The second equality comes from (\ref{mplanck}) with
$\omega_d$  given in Eq.~(\ref{omegad}) for $d> 2$, and
\begin{equation}
\label{omega2}
\omega_2=\frac{1}{2\,\pi}\log \frac{R \; \Lambda}{\pi}.
\end{equation}
Furthermore, for $d = 1$ we obtain
\begin{equation}
\label{m1}
\mathcal{M}_1^2= \frac{ M_{Pl}^2}{\mbar^2}\, \frac{\pi^2}{R^2}\ .
\end{equation}

Eq.~(\ref{mIR}) shows explicitly, for $d>2$,
the $R$ independent behavior of the mass, in close
relation to the result for
$\mathcal{M}_d$ in the $R\to\infty$ limit,
obtained in section 3, Eq.~(\ref{masad}).
Eq.~(\ref{m1})
shows that such a pole does not appear in the infrared
regime  for $d=1$. For $d=2$, the value of $\mathcal{M}_2^2$ is
given by
\begin{equation}
\label{m2c}
\mathcal{M}_2^2= \frac{ M_{Pl}^2}{\mbar^2}\,
\frac{1}{\omega_2 \; R^2}\ .
\end{equation}
and hence, from the expression of $\omega_2$ (\ref{omega2}), 
and excluding the unphysical case $\Lambda \gg M$, we obtain
that, unless $\mbar$ is of order $M_{Pl}$,
such a state will also be absent in the infrared
regime.

Since for $d=1$ the mass term (\ref{m1}) is, for $\mbar^2\ll M_{Pl}^2$, much
larger than $1/R$,  the second term in (\ref{GIRpoles}) corresponds, in the
IR region, to a contact term in the propagator. A similar effect
is obtained for $d = 2$.
For $d > 2$, instead,
the second term in (\ref{GIRpoles}) corresponds to a massive state
coupled with a strength $1/\mbar^2$. This state will appear in the 
physical spectrum of the theory
whenever $\mathcal{M}_d$ is in the energy range under consideration,
namely whenever $\mathcal{M}_d < 1/R$. This happens, 
for $M \simeq \Lambda$, when
\begin{equation}
\label{mbarIR}
\left(1-\frac{\mbar^2}{M_{Pl}^2}\right)^{\frac{1}{2}-\frac{1}{d}}
\mbar > \sqrt{\frac{1}{\omega_d}}\, \left(\frac{M_{Pl}}{M}\right)^{2/d} M
\simeq \sqrt{\frac{1}{\omega_d}}\, 10^{\,32/d}\, \left(\frac{\rm 1\, TeV}{M}
\right)^{2/d}\,M\ .
\end{equation}
As noticed above, the
inequality (\ref{mbarIR}) is only consistent with a coupling much stronger
than the gravitational one, i.e.
$\mbar\ll M_{Pl}$, for $d\geq 3$. In this case
the second term of (\ref{GIRpoles})
gives rise to a massive ``condensate'' with a mass
given by Eq.~(\ref{mIR}) and a coupling much stronger than the gravitational
one. This leads to a very novel phenomenon:
The presence of two gravitons in the spectrum, with masses lower
than the compactification scale $1/R$. One of the gravitons becomes
massless and mediates the regular gravitational interactions. The second
graviton, much more strongly coupled, could have a mass in the range
detectable by table-top gravitational experiments, as we
will see in the next section.

\subsection{The UV region}

In the UV region, $R\sqrt{s}\gg 1$, the Green functions for $d=1,2$ can be
taken directly from Eqs.~(\ref{5d}) and (\ref{6d}).
In particular for $d=1$ the Green function is,
\begin{equation}
\label{green1}
G_1(s)\simeq- \frac{R}{i\,2\, M_{Pl}^2\, \sqrt{s}+\mbar^2\, R\, s}\ .
\end{equation}
For $\sqrt{s}\gg r_c^{-1}$
($\sqrt{s}\ll r_c^{-1}$) the theory behaves four dimensional
(five dimensional), and the critical length is,
\begin{equation}
\label{pc5}
r_c\simeq \frac{\mbar^2\, R}{2\, M_{Pl}^2}\ .
\end{equation}

Assuming now that $M_{Pl}^2\simeq M^3\, R$, as discussed above,
we obtain from (\ref{pc5})
$r_c\simeq \mbar^2/2\, M^3$, the same critical length we got in section 3
for the $R\to\infty$ case. Moreover, for $d=1$ and $\bar{M}\ll M_{Pl}$,
$R\simeq 10^{15} {\rm cm} \times (1\ {\rm TeV}/M)^3$. Therefore,
unless $M > 10^5$ TeV, the theory  is ruled out by gravitational 
experiments, as in the ADD case.

For $d=2$ the Green function is,
\begin{equation}
\label{green2}
G_2(s)\simeq \frac{R^2\left[i\,\pi+\log\left(\Lambda^2/s\right)\right]}
{4\pi\, M_{Pl}^4-\mbar^2\, s\,R^2
\left[i\,\pi+\log\left(\Lambda^2/s\right)\right]}\ .
\end{equation}
Assuming again that $M_{Pl}^2\simeq M^4\, R^2$
we obtain from
(\ref{green2}) the critical length (\ref{rc6}) we got in the $R\to\infty$
case. Again, for values of $s$ close to $\mathcal{M}_2$ the Green function
(\ref{green2}) can be approximated by one describing a resonance of mass
$\mathcal{M}_2$ and width controlled by $\gamma_2$, Eq.~(\ref{gamma2}).
Due to a different behavior of the function $D_2(s)$
in the ultraviolet and the infrared regimes, the function $\omega_2$ used
in the computation of $\mathcal{M}_2$, Eq. (\ref{mIR}), in the
ultraviolet regime should take the form
\begin{equation}
\label{om2c}
\omega_2 = \frac{1}{2 \pi} \log\left(\frac{\Lambda}{\sqrt{s}}\right).
\end{equation}
and hence, for $\bar{M}\ll M_{Pl}$, $\mathcal{M}_2$ becomes 
independent of $R$ in this
regime, in agreement with the infinite radius case.

For $\mbar\ll M_{Pl}$, $\mathcal{M}_2$ tends to be large enough,
so that table-top experiments are not sensitive to
the state of mass $\mathcal{M}_2$. However, they are sensitive to ordinary
KK states with masses $\sim 1/R$. Since for $M = 1$ TeV,
the value of $R$, as fixed from (\ref{mplanck}), for  $\mbar\ll
M_{Pl}$, is  $R \simeq 1\,{\rm mm}$, and present results from gravitational
experiments are already sensitive to sub-millimeter distances~\cite{mm,mm2},
this case (similarly to ADD) demands values of $M$ larger than the
TeV scale.

Finally, for $d>2$ the $D_d(s)$ function has the expression,
\begin{equation}
\label{DUV}
D_d(s)=\frac{\Omega_d}{\pi^d}\,s^{\frac{d}{2}-1}
\left[i\,\frac{\pi}{2}+\frac{1}{d-2}\left(
\frac{\Lambda^2}{s}\right)^{\frac{d}{2}-1}\right]
\end{equation}
and the full Green function can be written as,
\begin{equation}
\label{green7}
G_d(s)\simeq -\frac{1}{M_{Pl}^2}\, \frac{1}{s}
-\frac{1}{\mbar^2}\left(1-\frac{\mbar^2}{M_{Pl}^2}\right)\,
\frac{s-\mathcal{M}_d^2+\gamma_d^2\, s-i\, \gamma_d\, \mathcal{M}_d^2}
{\left(\mathcal{M}_d^2-s\right)^2+\gamma_d^2\, s^2}
\end{equation}
which corresponds to the propagation of the normal massless mode plus a
massive (effective) resonance with a mass $\mathcal{M}_d$, given in
Eq.~(\ref{mIR}), and a width controlled by the function $\gamma_d(s)$
given by Eq.~(\ref{width}).

We observe that the pole structure of propagators in the UV-region
is, as expected, equivalent to  that of the theory in the $R\to\infty$ limit,
that was studied in section 3. The $R\to\infty$ limit can be safely taken
after the UV limit $R\sqrt{s}\gg 1$.
This is a consequence of the existence of a physical UV cutoff.

Finally notice that the result in the IR, Eq.~(\ref{GIRpoles}), can be formally
obtained  from (\ref{green7}) by taking $\gamma_d=0$. This behavior can be
understood from Eqs.~(\ref{oddM}) and (\ref{evenM}) because, in the region
$R\sqrt{s}\ll 1$ one can expand:
$\tanh^{-1}(\pi/R\sqrt{s})=-i\,\pi/2+\mathcal{O}(R\sqrt{s}/\pi)$ and
$\log(1-\pi^2/R^2 s)=i\pi+\log(\pi^2/R^2 s-1)$. Then the imaginary part of
the Green functions $D_d(s)$ cancels in this region, in agreement with the
results of the previous subsection.

\section{Phenomenological implications}

In this section we will study the modification of ADD phenomenology
by the presence of the brane correction term, assuming $\Lambda = M$. 
The relevance of the
presence of $\mbar$, and in particular of the state with mass $\mathcal{M}_d$,
does depend on the relative value of $\mathcal{M}_d$.
\begin{itemize}
\item
When $\mathcal{M}_d\simlt 1/R$, because of its weak coupling
the new state with mass $\mathcal{M}_d$ and
coupling to matter $\sim 1/\mbar^2$ is not expected to be detected in
collider experiments (see Eq. (\ref{mbarIR})) but, nevertheless has to
be considered in gravitational table-top experiments. The reason is that
its coupling can be much stronger than the Newton constant,
which governs the coupling of the ordinary KK states. On the other hand,
the coupling $\alpha$ of ordinary KK modes to matter is suppressed with
respect to the ADD coupling $\alpha_{\rm ADD}\simeq 1/M_{Pl}^2$ as,
$\alpha\simeq (\mathcal{M}_d/m)^4 \alpha_{\rm ADD}$, where $m$ is the
mass of the KK mode. Therefore, these states do not in general affect 
(or do it very mildly) table-top experiments.

\item
When $1/R < \mathcal{M}_d\simlt M$, and $d>2$, 
since the values of $R$ are $R \ll 1$ mm,
the new state is too heavy to affect the
gravitational table-top experiments.
However, depending on the value of $\mathcal{M}_d$ the coupling
to matter $\sim 1/\mbar^2$ may not be negligible and that state can
alter the collider phenomenology in a significant way. On the other hand,
KK modes which are much lighter than $\mathcal{M}_d$ couple to matter as in
the ADD scenario.
\end{itemize}

\subsection{Table-top gravitational phenomenology}

We will assume in this subsection that $\mathcal{M}_d\simlt 1/R$ and see
how the presence of the new state with mass $\mathcal{M}_d$ can affect the
table-top gravitational phenomenology.

The gravitational potential on the brane for distances
$r\simgt \mathcal{M}_d^{-1}$ can be written as:
\begin{equation}
\label{potencial}
V(r)=-G_N\frac{m_1\, m_2}{r}\left(1+\alpha\, e^{-r/\lambda}\right)
\end{equation}
where the parameters $\alpha$ and $\lambda$, as defined by
\begin{equation}
\label{parametros}
\alpha= \frac{M_{Pl}^2}{\mbar^2}-1,\quad
\lambda= \mathcal{M}_d^{-1}\ ,
\end{equation}
do have a mild dependence on the number of extra dimensions $d$ through
$\omega_d$ in (\ref{mIR}) and the consistency condition (\ref{mbarIR}).
Using then (\ref{parametros}) and (\ref{mIR}) we can write,
\begin{equation}
\label{relaciones}
\frac{(\alpha+1)^2}{\alpha}\,
\left(\frac{\lambda}{\rm mm}\right)^2\simeq \omega_d\, \left(\frac{\rm 1\, TeV}
{M}\right)^4
\end{equation}
while inequality (\ref{mbarIR}) and Eq.~(\ref{relaciones}),
imply for $\lambda$ the bounds
\begin{equation}
\label{cota}
10^{-16\, \frac{d-2}{d}}\left(\frac{\rm 1\, TeV}
{M}\right)^{\frac{2}{d}+1}\, <\frac{\lambda}{\rm mm}<
\frac{\sqrt{\omega_d}}{2}\left(\frac{\rm 1\, TeV}
{M}\right)^2 .
\end{equation}

\begin{figure}[ht]
\vspace{.5cm}
\centering
\epsfig{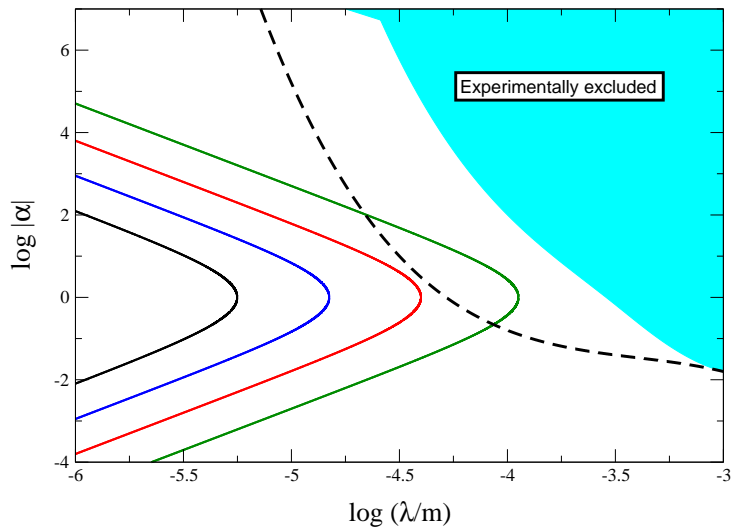}
\caption{\it Comparison with table-top experiments. Solid curves are
Eq.~(\ref{relaciones}) with $d=3,4,5,6$, from top to bottom, respectively,
and $M=1$ TeV.
The shaded region is excluded from present experimental data, Ref.~\cite{mm},
while the dashed curve is the projected experimental sensitivity of
Ref.~\cite{mm2}.}
\label{tabletop}
\end{figure}

In Fig.~\ref{tabletop} we plot Eq.~(\ref{relaciones}) for dimensions
$d=3,4,5$ and $6$, solid curves from top to bottom, respectively, and
$M=1$ TeV, and
show, in the $(|\alpha|,\lambda)$ plane the present excluded region
from table-top experiments, Ref.~\cite{mm}. The dashed line represents
the projected experimental sensitivity of Ref.~\cite{mm2}.
The crossing of solid and dashed curves will
impose some upper bounds on $\mbar$.

\subsection{Collider phenomenology}

We will assume in this subsection that $d \geq 2$ and 
$\mbar \simgt M$, so that
for the energies accessible at present and future high-energy collider
we can approximate the Green function as one where the large number of
KK modes behave like an ``effective'' resonance state, given by
Eqs.~(\ref{green7}), 
with mass $\mathcal{M}_d$ and width controlled by
$\gamma_d$, Eq.~(\ref{generalw}). 
In the $\mbar\to 0$ limit (i.e. $s\ll \mathcal{M}_d^2$) one
recovers the ADD (constant) result. However for values of $\mbar 
\simgt M$ there will be important modifications of ADD collider phenomenology,
both on KK graviton production and on virtual graviton mediated
processes, that will be briefly analyzed in this section.

\subsubsection{Production processes}

As it was already
emphasized in Ref.~\cite{ADD}, processes with KK graviton production
are an important signature of extra dimensions. They appear as missing-energy
events where a particle (photon $\gamma$, quark $q$ or gluon $g$) is
produced and no observable particle is balancing its transverse momentum.

Given the huge number of KK modes and, correspondingly, the smallness of
their mass differences, we can replace the production sum of KK modes by
a continuous integration, and write the differential cross-section for
inclusive graviton production (i.e. $f\bar f\to\gamma X$, $q\bar q\to g X$,
$q g\to q X$, $gg\to g X$) as
\begin{equation}
\label{produccion}
\frac{d\sigma}{dt}=\frac{1}{\pi}\int_0^s {\rm Im}\,\tilde{G}_d(m^2)
\frac{d\sigma_m}{dt}\, dm^2
\end{equation}
where ${\displaystyle \frac{d\sigma_m}{d\,t}}$
is the differential cross-section for production
of a single (canonically normalized)
KK mode with mass $m$
and a coupling strength to matter $1/M_{Pl}^2$, and
$\tilde{G}_d(m^2) = M_{Pl}^2\ G_d(m^2)$. The expression for
${\rm Im}\,\tilde{G}_d(m^2)$ is given by
\begin{equation}
\label{imaginario}
{\rm Im}\,\tilde{G}_d(m^2)=
\frac{M_{Pl}^2}{\mbar^2}\left(1-\frac{\mbar^2}{M_{Pl}^2}\right)\,
\frac{\gamma_d(m^2)\, \mathcal{M}_d^2}
{\left(\mathcal{M}_d^2-m^2\right)^2+\gamma_d^2(m^2)\, m^4}\ .
\end{equation}

From (\ref{imaginario}) we can see that for $m^2\ll \mathcal{M}_d^2$,
\begin{equation}
\label{lowKK}
{\rm Im}\, \tilde{G}_d(m^2)\simeq \frac{\pi}{2}\,\frac{\Omega_d}{\pi^d}\,
\frac{\left(m^2\right)^{d/2-1} M_{Pl}^2}{M^{2+d}}
={\rm Im}\,\tilde{G}_d(m^2)_{\rm ADD}\ .
\end{equation}
In this way the KK modes which are much lighter than $\mathcal{M}_d$
couple to matter as in the ADD scenario. Therefore, if
$\sqrt{s} < \mathcal{M}_d$, the differential cross section for the reaction
$e^+ e^- \to \gamma X$ is
\begin{equation}
\left.
\frac{d \sigma}{d \cos\theta} \right|_{\sqrt{s} < \mathcal{M}_d}
\simeq \frac{\alpha}{M^2}
\left( \frac{\sqrt{s}}{M} \right)^{d} (1 + \cos^2\theta)
\end{equation}
where $\alpha$ is the electromagnetic constant and $\theta$ the scattering
angle in the center-of-mass system.

On the other hand, for
$m^2\gg \mathcal{M}_d^2$ the imaginary part of the Green function behaves as,
\begin{equation}
\label{highKK}
{\rm Im}\,\tilde{G}_d(m^2)\simeq \left(
\frac{\mathcal{M}_d}{m}\right)^4\,{\rm Im}\,\tilde{G}_d(m^2)_{\rm ADD}
\end{equation}
and the corresponding KK modes couple to matter as
$\sim\mathcal{M}^4_d/m^4$ times the ordinary Newton constant.

To estimate the integral in (\ref{produccion}) we have found, 
for $\sqrt{s}\simgt \mathcal{M}_d$, to a good accuracy,
\begin{equation}
\label{prodG}
\frac{d\sigma}{dt}\simeq \frac{M_{Pl}^2}{\mbar^2}\,
\left(1-\frac{\mbar^2}{M_{Pl}^2}
\right) \frac{d\sigma_m(\mathcal{M}^2_d)}{dt}
\end{equation}
where we have used the smallness of $\gamma_d$ and the property
$\lim_{\varepsilon\to 0}\varepsilon/(x^2+\varepsilon^2)=\pi\delta(x)$.
In this way, the collective effect of all KK modes behaves as the production
of a single sharp resonance of mass $\mathcal{M}_d$ and coupling $1/\mbar^2$.
This signature is very distinct from ADD. For example,
for $e^+ e^-\to \gamma X$,
and neglecting terms of $\mathcal{O}(\mathcal{M}_d^2/s)$, the differential
cross section (\ref{prodG}) can be written as~\cite{colliders}
\begin{equation}
\label{ee}
\frac{d\sigma}{d\, \cos\theta}\simeq \frac{\pi\alpha}{\mbar^2}(1+\cos^2\theta)
\ .
\end{equation}
The modes with mass smaller and larger
than $\mathcal {M}_d$ give a total contribution proportional to
$1/M^2 \left(\mathcal{M}_d/M\right)^d$ and $1/\bar{M}^2
\left(s/M^2 \right)^{d/2-1} \left(\mathcal{M}^2_d/s\right)$, respectively.
These contributions
are sub-dominant for $d >2$, while for $d = 2$ can be at most of
order of the contribution of the collective mode of mass
$\mathcal{M}_d$, Eq.~(\ref{ee}).
In this way the present bounds from
this process at LEP2 in ADD can be easily translated into the lower bound
$\mbar\simgt 5$ TeV whenever $\mathcal{M}_d$ is kinematically accessible
to this machine. If $\mathcal{M}_d$ is not energetically accessible, the
bound on $\mbar$ disappears and it is replaced by a bound on $M$ similar to
that obtained in the ADD case~\cite{colliders}.
Of course much stronger bounds will be obtained from
future accelerators.

\subsubsection{Virtual exchange}

The effective resonance of mass $\mathcal{M}_d$ can also contribute to
physical processes by a single virtual exchange in the s-channel, along
with the exchange of
other standard model particles. It can contribute sizeably to the
cross-sections if the center-of-mass energy $s$ is close to $\mathcal{M}_d$.
In that case its contribution to the cross-sections is proportional to
\begin{figure}[htb]
\vspace{.5cm}
\centering
\epsfig{file=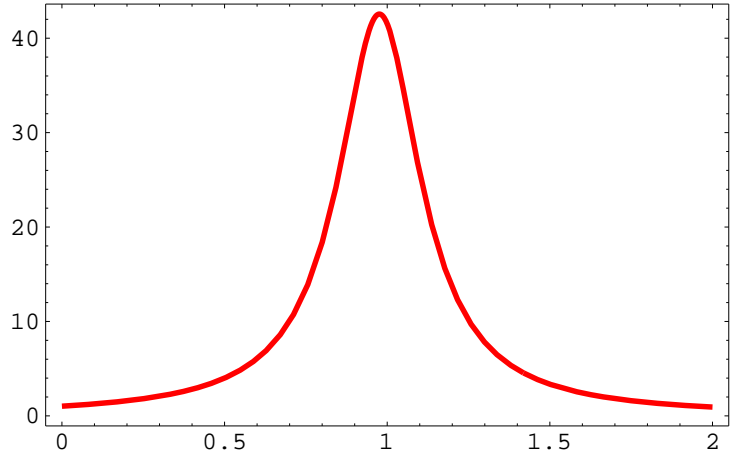,width=.6\linewidth}
\caption{\it Plot of the function $f(s/\mathcal{M}_d)$ for $d=3$ and
$\mathcal{M}_d=0.1\, M$.}
\label{colliders}
\end{figure}
\begin{equation}
\label{exchange}
\frac{1}{\mbar^4\mathcal{M}_d^4}\,f(s/\mathcal{M}_d^2)
\end{equation}
\noindent
with
\begin{equation}
f(s/\mathcal{M}_d^2)=\mathcal{M}_d^4\
\frac{\left(s-\mathcal{M}_d^2+\gamma_d^2(s)\, s\right)^2+
\gamma_d^2(s)\mathcal{M}_d^4}{\left[\left(\mathcal{M}_d^2-s\right)^2+
\gamma_d^2(s)\, s^2\right]^2}\ .
\end{equation}
At the resonance peak there is an enhancement, with respect to the (constant)
ADD contribution, as
$$\left(\frac{\mbar^2}{M^2}\right)^{d-2}$$
while the distribution width 
$$\sim \frac{\pi}{2}(d-2)\left(\frac{\mathcal{M}_d}{M}\right)^{d-2}$$
is essentially governed by the ratio $\mathcal{M}_d/M$.
An example of the resonance distribution is given in Fig.~\ref{colliders},
where the function $f(s/\mathcal{M}_d)$ is plotted  for the case
$\mathcal{M}_d/M=0.1$ and $d=3$.

\section{Conclusions}

This paper deals with the effect of kinetic brane gravitational corrections on
extra dimensional scenarios. These corrections can be induced by interactions
of bulk gravitons with matter localized on the brane and
their size is not protected by any four dimensional symmetry acting on the
brane. To avoid the complication inherent to the tensor structure of the
higher dimensional graviton we have worked out a prototype model with a
simple scalar field propagating in the bulk of the extra dimensions.
This work was partly motivated by the recent and interesting claim that
for the case of
two (or more) infinite size flat 
extra dimensions kinetic brane corrections
trigger higher dimensional gravitons to be localized on the brane. Moreover,
for finite size radius we also expect kinetic brane gravitational
corrections to substantially modify the usual picture initially introduced
by Arkani-Hamed, Dimopoulos and Dvali as an alternative solution to the
hierarchy problem and, in particular, its implications on collider
phenomenology and table-top gravitational experiments.

We have endowed the theory with both an IR and an UV regularization.
The IR regularization is implemented by a $d$ dimensional torus with a common
radius $R$. The infinite size can then be reached in the $R\to\infty$ limit.
The UV regularization is provided by a physical UV cutoff $\Lambda$.
Since the higher dimensional gravitational theory is an effective one, with
a physical cutoff of the order of the higher dimensional Planck scale, $M$,
the cutoff $\Lambda$ is restricted to be $\sim M$, or even less if the
brane is fat, a possibility that, for simplicity, we are not considering.

For the case of infinite size extra dimensions, $R\to\infty$, were we
allowed to take the $\Lambda\to\infty$ limit, we would recover the results
of Refs.~\cite{porrati,dvali}. However, as $\Lambda\simlt M$, as we
pointed out above, our results for infinite size extra dimensions
do not lead to the existence of tensor gravity localized on the brane but,
instead, lead in general to a massive graviton. To summarize, in the
absence of a quantum field theory of gravity in higher dimensions, the
existence of a physical UV cutoff prevents localization of tensor gravity on
the brane in the presence of infinite size extra dimensions.

For the case of finite extra dimensions, kinetic brane gravitational
corrections induce deep modifications on the ADD scenario.
In particular the emergence of a ``collective'' state (made out of an
``infinite'' number of KK modes) with a mass $\mathcal{M}_d \sim
M^2/\mbar$ and an effective coupling $G_{\rm eff}\sim 1/\mbar^2$, where
$\mbar$ is the brane correction. For $\mathcal{M}_d \simlt 1/R$, this
state can be detected in table-top gravitational experiments, while ADD
collider phenomenology is deeply modified since KK gravitons become
unobservable in high-energy colliders. On the other hand, for
$1/R\simlt \mathcal{M}_d \simlt M$, 
table-top gravitational experiments are blind to the
new state and so table-top gravitational phenomenology
remains essentially unchanged. On the other hand the new resonance can be
produced on-shell at high-energy colliders and then also modify the ADD
phenomenology.

\section*{Acknowledgments}
This work has been supported in part
by the US Department of Energy, High Energy Physics Division,
under Contracts DE-AC02-76CHO3000 and W-31-109-Eng-38,
by CICYT, Spain, under contract AEN98-0816, by EU under
RTN contracts HPRN-CT-2000-00152 and HPRN-CT-2000-00148, and by the
Polish State Committee for Scientific Research, grant KBN 2 P03B 060 18
(2000-01). The work of AD was supported by the Spanish Education Office
(MEC). MC and CEMW wish to thank G.~Dvali and V.A.~Rubakov for
useful discussions.

\appendix

\section{Appendix}
The integral (\ref{integral1}) can be written in general for
even and odd values of $d$.

\begin{itemize}
\item
For $d$ odd:

\begin{align}
&
D_d(p,0)=\frac{1}{p^2 R^d} \left\{1+\Omega_d \frac{R^d p^d}{\pi^d}
\left[(-1)^{(d-1)/2}\left(\arctan\left(
\frac{\Lambda}{p}\right)
-\arctan\left(\frac{\pi}{Rp}\right)\right)
\phantom{\frac{1^{1^{1}}}{1^{1^{1}}}}\right.\right.\nonumber\\
&+ \left.\left.\sum_{n=1}^{(d-1)/2}
\frac{(-1)^{n-1}}{d-2n}\left(\left(\frac{\Lambda}{p}\right)^{d-2n}-
\left(\frac{\pi}{Rp}\right)^{d-2n}\right)\right]\right\} \ .
\label{odd}
\end{align}
\item
For $d$ even:

\begin{align}
&
D_d(p,0)=\frac{1}{p^2 R^d} \left\{1+\Omega_d
\frac{R^d p^d}{\pi^d}\left[(-1)^{d/2-1}\frac{1}{2}
\left(\log\left(1+\frac{\Lambda^2}{p^2}\right)-
\log\left(1+\frac{\pi^2}{R^2\,p^2}\right)
\right)
\phantom{\frac{1^{1^{1}}}{1^{1^{1}}}}\right.\right.\nonumber\\
&+ \left.\left.\sum_{n=1}^{d/2-1}\frac{(-1)^{n-1}}{d-2n}
\left(\left(\frac{\Lambda}{p}\right)^{d-2n}-
\left(\frac{\pi}{Rp}\right)^{d-2n}\right)\right]\right\}\ .
\label{even}
\end{align}
\end{itemize}

The analytic continuation of the function $D_d(p,0)$ into Minkowski
space-time, $D_d(s)$, can be done by simply replacing in (\ref{odd}) and
(\ref{even}) $p\to -i\sqrt{s}$ and using the property,
$\arctan(ix)=i\tanh^{-1}(x)$. The result can be written as:

\begin{itemize}
\item
For $d$ odd:

\begin{align}
&
D_d(s)=\frac{1}{s\, R^d} \left\{-1+\Omega_d \frac{R^d}{\pi^d}\, s^{d/2}
\left[i\,\frac{\pi}{2}-\tanh^{-1}\left(
\frac{\sqrt{s}}{\Lambda}\right)
+\tanh^{-1}\left(\frac{\pi}{R\,\sqrt{s}}\right)
\phantom{\frac{1^{1^{1}}}{1^{1^{1}}}}\right.\right.\nonumber\\
&+ \left.\left.\sum_{n=1}^{(d-1)/2}
\frac{1}{d-2n}\left(\left(\frac{\Lambda^2}{s}\right)^{\frac{d}{2}-n}-
\left(\frac{\pi^2}{R^2\,s}\right)^{\frac{d}{2}-n}\right)\right]\right\} \ .
\label{oddM}
\end{align}
\item
For $d$ even:

\begin{align}
&
D_d(s)=\frac{1}{s\, R^d} \left\{-1+\Omega_d \frac{R^d}{\pi^d}\, s^{d/2}
\left[i\,\frac{\pi}{2}+\frac{1}{2}\log\left(\frac{\Lambda^2}{s}-1\right)
-\frac{1}{2}\log\left(1-\frac{\pi^2}{R^2\,s}\right)
\phantom{\frac{1^{1^{1}}}{1^{1^{1}}}}\right.\right.\nonumber\\
&+ \left.\left.\sum_{n=1}^{\frac{d}{2}-1}
\frac{1}{d-2n}\left(\left(\frac{\Lambda^2}{s}\right)^{\frac{d}{2}-n}-
\left(\frac{\pi^2}{R^2\,s}\right)^{\frac{d}{2}-n}\right)\right]\right\} \ .
\label{evenM}
\end{align}
\end{itemize}
The imaginary part proportional to $\pi/2$ in Eqs.~(\ref{oddM}) and
(\ref{evenM}) come from the resonant production of a single KK mode with
square mass equal to $s$ while the real part comes from the summation
over non-resonant states.

Substituting the functions $D_d(p,0)$, as given by (\ref{odd}) or (\ref{even}),
and $D_d(s)$,as given by (\ref{oddM}) or (\ref{evenM}),
into Eq.~(\ref{solution1}) provides the Green function $G_d(p,0)$, and
its analytic continuation into the Minkowski space-time $G_d(s)$, for
the bulk scalars.

\end{document}